\documentclass[runningheads]{llncs}
\pdfoutput=1

\usepackage{cite}
\usepackage{amsmath,amssymb,amsfonts}
\usepackage{microtype}
\usepackage{algorithmic}
\usepackage{graphicx}
\usepackage{textcomp}
\usepackage{bmpsize}
\usepackage{xcolor}
\usepackage{lipsum}
\usepackage[colorlinks=true,urlcolor=black]{hyperref}
\def\BibTeX{{\rm B\kern-.05em{\sc i\kern-.025em b}\kern-.08em
    T\kern-.1667em\lower.7ex\hbox{E}\kern-.125emX}}
\begin{document}

\title{"It’s Not Something We Have Talked to Our Team About": Results From a Preliminary Investigation of Cybersecurity Challenges in Denmark}

\author{Camilla Nadja Fleron \and
Jonas Kofod Jørgensen \and
Oksana Kulyk \and 
Elda Paja}
\institute{IT University of Copenhagen, Denmark  \\
\email{\{cafl,jkoj,okku,elpa\}@itu.dk}
}

\maketitle
\begin{abstract}
Although Denmark is reportedly one of the most digitised countries in Europe, IT security in Danish companies has not followed along. To shed light into the challenges that companies experience with implementing IT security, we conducted a preliminary study running semi-structured interviews with four employees from four different companies, asking about their IT security and what they need to reduce risks of cyber threats. Our results show that companies are lacking fundamental security protection and are in need of guidance and tools to help them implementing basic security practices, while raising awareness of cyber threats. Based on our findings and with the inspiration of the latest reports and international security standards, we discuss steps towards further investigation towards developing a framework targeting SMEs that want to adopt straightforward and actionable IT security guidance. 
\end{abstract}

\section{Introduction}

With the rise of internet usage as a means of doing business and gaining access to large amounts of data, organisations all over the world increasingly depend on IT systems \cite{b1_ds}. This dependence is particularly prominent in countries with high level of digitisation, which according to the European Commission's Digital Economy and Society Index (DESI) of 2019 \cite{b2_desi2019} include Denmark, Sweden, Finland, and the Netherlands. With the high level of digitisation, however, protecting against cybersecurity risks becomes of crucial importance. 

However, IT security in Danish companies has not followed the high digitisation degree. According to a survey conducted by Monitor Deloitte in 2018 \cite{b3_monitor}, 22\% of Danish small and medium-sized enterprises (SMEs) do not have even the most basic IT security implemented (including e.g. setting up a firewall or implementing access control measures), 39\% of SMEs have an IT security level that is considered inadequate to their risk profile, while 70\% of Danish SMEs end up outsourcing their IT security. The research suggests that the companies often do not have the necessary competences in the field to be able to establish and operate the required security measures. While there exist plenty of standards \cite{b4_ISOreq}\cite{b5_NISTcore}, frameworks\cite{b6_cyberforsvar}, badges\cite{b7_cyberEssentials, b8_maerkning} and internet tools\cite{b9_sikkerhedstjekket}, which may assist with securing one's organisation, navigating them can be overwhelming. As such, some of the tools may not fit the risk profile of the SME, and some may be too complicated for companies with limited resources and security expertise. 

The purpose of this work is that of getting a deeper understanding into the challenges faced by Danish SMEs in establishing and managing security practices. As such, we conduct exploratory semi-structured interviews, investigating (i) the \textit{current status of cybersecurity} in companies, (ii) the company \textit{security culture}, and (iii) \textit{the kind of help the companies seek to better manage their security}. We report on findings, and discuss our research plans to tackle the needs identified by this preliminary study. 

\section{Study Methodology} 
\label{sec:method}

\label{sec:rqs}
In our study, we explore the following research questions: 
\begin{itemize}
    \item RQ1: What is the level of awareness about security in companies?
    \item RQ2: What is the companies level of security protection?
    \item RQ3: What help might the companies need concerning security?
\end{itemize} 

The interview questions are inspired by the work of Beautement et al. \cite{b10_beautement}, which investigated the relationship between organisational security policies and employees' behaviour toward these. The Monitor Deliotte study \cite{b3_monitor} provided us already with an indication of SMEs current state of security protection, and at the same time with examples of interview questions. 

Previously to running the interviews, the interview guide was tested in a pilot study. After it was finalised, the interview guide was sent to the respondents before the actual interview in order to make the respondents confident about the agenda and what kind of questions they were going to answer. Since the interview focused on IT security practices of the organisation, which might be considered confidential information to the company, the participants were also informed in advance that the findings will be reported without enabling any link to the respondent's identity or their company, and that they would have a chance to review the interview transcripts before further processing.  The interview guide is provided in the Appendix (translated from Danish). 

The study was conducted in respect of participants privacy and anonymity. To this end, all collected recordings have been destroyed after transcription. Names of people and companies have been anonymised. To further respect participants privacy, as promised to the participants before the interview, each participant was offered to review the transcript. The interviews were recorded and transcribed immediately after taking place. The transcripts were analysed via content analysis \cite{b11_hancock}, with regards to the research questions outlined above.

\section{Results}
\label{sec:results}
We interviewed four people from four different companies, whose positions ranged from part-time worker to CEO. All participants had a university degree and at least one-year working experience\footnote{We do not report on participants demographics for the sake of  anonymity.}. Due to the corona pandemic we were not able to conduct all the interviews face-to-face. Hence, two of the interviews  took place in person before the society shut down, while the last two were conducted remotely. The duration of the interviews  ranged between 20 and 30 minutes. We provide more details of our findings in the remainder of the section, where we look at the participants' answers to the specific research questions.

\subsection{RQ1: Awareness About Security}

In order to investigate the awareness about security in participants' companies, we look at the following questions: (1) whether they talk about IT security in their company, (2) whether there is security training in their company, and (3) whether there is a response plan, i.e. employees are aware of what they need to do in case of a security incident.

\subsubsection{Talking about IT-security.}
All respondents confirmed that they do not talk about security in their companies. Such attitude, furthermore, was expressed by one of the participants being the CEO of the company, hinting at a general problem with possible lack of involvement from the top-level. 

\subsubsection{Security Trainings.}

None of the companies had a formalised approach to the employees  awareness and knowledge in relation to IT security training. 

\subsubsection{Response Plan.}

Only one out of four participants knew that there was a response plan in the company that outlined what should be done in case of an IT-related incident. However, the participant did not know what the plan entailed, only that such a plan existed in the company. The participant furthermore mentioned that the response plan had not been updated recently and that it was created some years back and might need a revision. 

\subsection{RQ2: Security Measures}

We investigate security measures asking about (1) security policies in the company in general, (2) password policies specifically, and (3) history of security breaches and implemented countermeasures in the company.

\subsubsection{Security Policies.}

When asked about security guidelines in their company, only one participant was able to recall and describe them in details, mentioning policies such as locking the screen when one is away from one's computer, using tools like hard disk encryption, firewalls and MFA, as well as guidelines on choosing strong passwords and securely storing one's IT equipment. Another participant mentioned having to sign an IT security policy paper, but was not sure about the details contained therein. The remaining two interviewed companies had no security policy in place.

\subsubsection{Password Policies.} 

Password management is a hard discipline to master on the one hand, while it is crucial, on the other, since passwords provide an essential layer of protection. In this regard, participants reported difficulties in password management, resulting in workarounds such as using post-its for writing the passwords down. Further issues with passwords included non-optimal password policies, such as mandatory password change \cite{b13_ftc}.

\subsubsection{Security Breaches.}

We asked the participants if their company had experienced a cyber attack. Two of the participants did not know if their company had been attacked. The other two explained how they experienced cyber attacks, elaborately. When asked about measures taken after the security breach, the participants, however, expressed doubt whether breaches could be prevented in the future. In one of the cases, the company attempted to take additional security measures after the breach (namely, using 2 Factor Authentication), but the participant voiced concern that these measures were not followed through.

\subsection{RQ3: Helping Improve Security}

We asked the participants to think of anything they would need in their work to be better protected from a security perspective. The question was asked to get an understanding of the needs SMEs have to tackle the challenges concerning managing IT security. One participant was not sure about what was needed, but the other three formulated precise demands. Two of the participants were mentioning an \textit{awareness guide} of basic IT security or a brief explanation of common missing security practices and how to overcome them. Another participant emphasised the importance for such a guide to be \textit{low cost}. The participant from a company which had well-established security culture furthermore touched upon the need for a practice to \textit{regularly revisit the security measures} and mentioned a demand for a yearly plan that allowed to do such check-ups. 

\section{Discussion and Future Work}
\label{sec:future}

Our study emphasizes the lack of security practices and adequate security protection among Danish SMEs companies, while at the same time our initial results highlight the need in these companies for having at least some essential basic security guidance. Additionally, the study findings suggest that there is a need for guidance for SMEs so that they can integrate the required basic security essentials into their business practices. The need for such a guidance is pressing, but any efforts should consider that SMEs in particular might have scarce resources both in terms of budget for security and with respect to human expertise. While plenty of frameworks and standards have been developed, they might nonetheless be underutilised by companies. Closing this gap would require assessing such tools and making efforts to improve their usability and ease of use for increasing adoption, or developing new tools that, in addition to providing guidelines for effective security management, can actually be used by smaller companies who have limited expertise or resources. Ideally, such a solution refines existing standards and solutions to avoid reinventing the wheel. We advocate the need for following a layered approach to account for the various company budgets and in house security expertise. In this way, a basic layer would consist of steps that can be implemented without much difficulties by most of the companies, such as setting up a firewall or enabling automatic updates of systems. Examples of such measures can be found within the \textit{Cyber Essential} badge, claiming that the proposed measures could prevent up to 80\% of all cyberattacks \cite{b14_itgover}. Further layers of the framework would include steps that involve more advanced measures, such as regular training and awareness campaigns for the employees, steps that can be taken by companies who are ready to dedicate more resources into security. As with most of the support tools, a trade-off between providing succinct, understandable and actionable guidance, and ensuring that this guidance is comprehensive enough, needs to be made. The scope of both basic and more advanced layers would therefore have to be iteratively refined, involving conversations with both security and privacy experts and the companies themselves.

Based on the study findings, our research plan considers future work on (i) developing a security guide, and (ii) developing a layered security model for SMEs. Specifically on (i), we will focus on the ways to put the identified steps of the security framework into an understandable guide. For this, coming up with concrete guidance or instructions would be the next logical development of this work, making use of the already available resources by the information security community or official governmental agencies. Deciding on the concrete form of the guide, such as its medium (e.g. implementing it as a website, app or paper information materials) and the target audience (i.e. whether it should focus on management or aim to target other employees as well) would be a further step. Following the human-centered security by design approach, the design and development of the guide would need to be done in close collaboration with the companies who would need such support. Involving companies would also be a crucial component of the evaluation of the guide with practitioners. This could include designing an approach to test the usability and user experience of the guide, as well as the effect of using the guide on adoption of secure behaviors.

Point (ii) on the other hand considers future investigation on the extent to which the guide would need to be personalised. The layered security model entails a certain degree of personalisation, as it enables companies to decide how many steps of the framework they would be ready to implement. Depending on the company profile (e.g., whether it handles a lot of sensitive data, whether the employees work on-site or remotely, how much technical expertise do the employees have, etc.), one might need to introduce new dimensions and layers.

\section*{Appendix}

We provide the preliminary information that was sent to the participants before the interview and the questions asked during the interview below, both translated from Danish.

\subsection{Preliminary information}

\subsubsection{Interview:}

\begin{itemize}
    \item Concerning the IT-security state in small and medium size businesses
    \item The interview will take about 20-40 mins.
\end{itemize}

\subsubsection{Why we want to interview you:} 

\begin{itemize}
    \item We want to interview you to get an idea of how we could help you and your company be better protected against cyber-attacks.
    \item We would like to know how your company are handling IT-security, which policies and practices they have, and how you go around them.
    \item Get an idea of which kind of IT-security product that would be beneficial for your company.
\end{itemize}

\subsubsection{Anonymity:}

\begin{itemize}
    \item Confidentiality is highly prioritized!
    \item Everything that you say will be 100\% anonymous.
    \item The company you are working in will not be mentioned anywhere.
    \item We delete the audio records of the interview right after we have made the transcription.
    \item The transcription will mention you like this e.g: \emph{P4 "We are using our own devices at work"}
    \item We will send you the transcription of the interview which you can review and edit if wanted.
\end{itemize}

\subsubsection{The agenda for the interview:}

\begin{itemize}
    \item About you and your company
    \item IT-security policies and practices
    \item The attitudes to the company’s security policy
    \item What kind of IT-security help you might need
\end{itemize}

\subsection{Interview questions}

\subsubsection{Brief explanation before the interview starts:}

\begin{itemize}
\item Everything that you say will be 100\% anonymous
\item We will not mention your company anywhere
\item The reason why we are doing the interview is to gain information about IT-security
challenges and procedures in order to get an idea of which product that would suit you and
your company.
\item You don’t have to answer all questions -- you can always drop out of the interview
\end{itemize}

\subsubsection{About you and your company:}

\begin{itemize}
\item Tell us a little about the company (You don't have to mention it by name)
\item How many employees are you?
\item How long has the business been around?
\item What responsibilities and tasks do you have?
\end{itemize}

\subsubsection{IT-security issues}

\begin{itemize}
    \item How do you rate your general knowledge of IT security?
    \item Is IT security something you talk about at work? Why / why not?
    \item Have you had any IT security course or similar here at the company? Which ones?
    \item (Follow up if they don't understand the issue) ...like phishing training, best practices for
staying protected, privacy or similar?
    \item Do you have any defined IT security policies in the company?
    \item (If unsure supply with some of these):
    \begin{itemize}
        \item Passwords: No reuse of codes, no physical storage of codes.
        \item Access Control: Does everyone have access to everything? 2FA?
        \item System updates.
        \item Firewalls/antivirus programs
        \item Data sharing/data storage: backup and encryption.
        \item Physical security: How to get into your office?)
    \end{itemize}
    \item Are you personally concerned about a cyber-attack? Why / why not?
    \item Are you worried that your company will be exposed to a cyber-attack? Why / why not?
    \item Do you use your private devices at work or have a work computer and work mobile?
    \item Do you have a password on your laptop? (Follow-up: Why?)
    \item Do you reuse your passwords? Why / why not?
    \item Do you use passwords to access confidential or sensitive information in the company?
    \item Do you think a hacker would be interested in your company? (Why / why not?)
    \item (Give examples if they feel uncertain about what is meant) Customer information, payment information, confidential information, social security numbers, logins, passwords, money, reputation, web shop, databases
    \item Has the company faced a cyber-attack? If so, what kind of attack? What was stolen? Changes made after the trial?
    \item Does your company have a plan for what you do as a company if you are attacked?
    \item What kind of help could you think of that could be of use to make your work more secure?
    \item What would you think could help you as a company to be better protected?
    \item If there was a product that could help you with IT security in the company, what should it do?
\end{itemize}

\subsubsection{The attitudes to the company's security policy}

\begin{itemize}
    \item IT security and productivity are always a delicate balance, as more IT security in most cases lowers productivity. Can you think of a scenario where the security policies were too cumbersome and by that reduced your productivity, and you chose not to follow the security policy?
    \item Do you have any last comments?
\end{itemize}

\end{document}